\documentclass[final,5p,times,twocolumn]{elsarticle}

\usepackage{dsfont}
\usepackage[T1]{fontenc}
\usepackage[english]{babel}
\usepackage[latin1]{inputenc}
\usepackage{epsfig}
\usepackage{graphicx}
\usepackage{dcolumn}
\usepackage{bm}
\usepackage{xcolor}
\newcommand {\be}{\begin{equation}}
\newcommand {\ee}{\end{equation}}
\newcommand {\ba}{\begin{eqnarray}}
\newcommand {\ea}{\end{eqnarray}}

\usepackage{amsmath}  
\usepackage{amsfonts} 
\usepackage{nicefrac}
\usepackage{hyperref}
\usepackage[normalem]{ulem}
\usepackage{cancel}
\usepackage{dblfloatfix}
\usepackage{float}

\journal{Physica A: Statistical Mechanics and its Applications}

\begin{document}

\begin{frontmatter}

\title{Entropy-production fluctuation theorem for a generalized Langevin \\particle 
in crossed electric and magnetic fields}

\author[uami]{L. C. Gonz\'alez-Morales}
\author[uami]{I. P\'erez Castillo}
\author[uami]{J. I. Jim\'enez-Aquino}

\address[uami]{Departamento de F\'isica, Universidad Aut\'onoma Metropolitana-Iztapalapa, C.P. 09310, CDMX, M\'exico.}

\begin{abstract}
We study fluctuations of entropy production for a charged Brownian particle confined in a harmonic trap and driven out of equilibrium by crossed electric and magnetic fields. The magnetic field is constant and perpendicular to the plane of motion, while the electric field is time dependent and provides the driving. The non-Markovian dynamics is modeled by a generalized Langevin equation with memory and Gaussian noise. This setting represents a charged Brownian degree of freedom in a structured bath, where delayed friction modifies relaxation while the magnetic field couples the transverse coordinates. Using the exact solution of this linear dynamics, we obtain the time-dependent Gaussian phase-space probability density and from it compute the trajectory-dependent total entropy production. For two solvable driving protocols---direct forcing by a prescribed time-dependent electric force and dragging of the harmonic trap center---we prove analytically that the total entropy production obeys a detailed fluctuation theorem.
\end{abstract}

\end{frontmatter}

\section{Introduction}
Fluctuation theorems (FTs) are exact relations, first established in the pioneering works of the 1980s and 1990s \cite{Boshkov1981,Evans1993,Gallavoti1995,Jarzynski1997,Crooks1999,Kurchan1999}, that quantify irreversibility in nonequilibrium statistical physics. They sharpen the second law for small systems observed over finite times by relating the probability of a forward trajectory to that of an appropriate backward (time-reversed) trajectory. Beyond their original formulations, FTs have been developed for a broad family of thermodynamic observables, including work, heat, power, and entropy production \cite{vanZon2004,Noh2012,Jimenez2013,Bochkov2013,Lemos2018,Rao2018,Park2023,Semeraro2024}. Current activity continues to extend both their scope and interpretations, ranging from fluctuation relations for  systems in a constant magnetic field \cite{Coretti2020}, to topological formulations and feedback-controlled settings (Maxwell-demon-type scenarios) to connections with learning dynamics and even simplified game-like models \cite{Sidajaya2025,Hack2023,Mahault2022,Zeng2021}. Fluctuation relations have found applications in biophysical systems \cite{Hayashi2018}, in active matter \cite{Chaki2018,Goswami2018,Narinder2021,Argun2016}, and have also been extended to the quantum regime \cite{Funo2018,Binder2018}, and quantum curved spacetime \cite{Basso2025}.

In the present problem, the generalized Langevin equation provides a natural effective description of a Brownian degree of freedom coupled to bath variables that have been eliminated from the explicit dynamics. The memory kernel represents delayed friction, while the colored noise is constrained by the fluctuation--dissipation theorem. Such descriptions arise for trapped particles or ions in structured environments, viscoelastic media, hydrodynamic-memory regimes, and related anomalous-transport settings in which a purely Markovian white-noise Langevin equation does not capture the observed relaxation. The broader anomalous-diffusion literature illustrates how memory, initial preparation, confinement, ageing, and nonergodicity can influence relaxation and time-averaged observables \cite{Metzler2014,Cherstvy2018OU,Cherstvy2021Diffusivity,Wang2022Resetting}. In stochastic thermodynamics, related non-Markovian effects have also been studied in GLE-type systems \cite{Mai2007,Ohkuma2007,Jimenez2015,Ghosh2017,Cockrell2022}. The Lorentz force adds a distinct physical ingredient: it does no mechanical work by itself, but it rotates the response in the plane perpendicular to the magnetic field and changes the structure of probability currents in Brownian systems with magnetic forcing \cite{Rosinberg2015,Abdoli2020}.

In 2005, Seifert \cite{Seifert2005} extended the fluctuation-theorem framework to entropy production and proved a detailed fluctuation theorem (DFT) in a nonequilibrium steady state over a finite time interval, $P(\Delta s_{\rm tot})/P(-\Delta s_{\rm tot})=e^{\Delta s_{\rm tot}}$, where $P(\Delta s_{\rm tot})$ is the probability of observing a total entropy production $\Delta s_{\rm tot}$ along a forward trajectory and $P(-\Delta s_{\rm tot})$ is the corresponding probability for the backward trajectory. The total entropy production $\Delta s_{\rm tot}$ includes both the change in system entropy (the Brownian particle) and the entropy change of the surrounding bath. The associated integral fluctuation theorem (IFT),
$\langle e^{-\Delta s_{\rm tot}}\rangle=1$, follows directly from the DFT.
Four years later, the DFT was extended to transient processes initialized from an equilibrium canonical distribution \cite{Saha2009}. Subsequently, the transient DFT was established for an electrically charged Brownian particle in the presence of crossed electric and magnetic fields \cite{Jimenez2010}. The analyses in \cite{Seifert2005,Saha2009,Jimenez2010} were formulated for Markovian stochastic dynamics described by a standard Langevin equation with additive Gaussian white noise.

On the other hand, fluctuation relations such as Crooks' theorem and Jarzynski's equality have also been extended to non-Markovian dynamics described by a generalized Langevin equation (GLE) with additive colored noise \cite{Mai2007,Ohkuma2007,Jimenez2015}. In particular, Ref.~\cite{Jimenez2015} investigated the transient work fluctuation theorem for a charged Brownian particle in crossed electric and magnetic fields within a GLE framework with a time-dependent friction memory kernel. The corresponding detailed fluctuation theorem for the total entropy production in the non-Markovian setting was addressed in Ref.~\cite{Ghosh2017}, where the analysis relies on a Fokker-Planck description for the joint probability density $P(x,w,t)$ of the particle position $x$, the thermodynamic work $w$, and time $t$.
 
To the best of our knowledge, an explicit demonstration of the DFT for the total entropy production of a charged Brownian particle in crossed electric and magnetic fields in the non-Markovian (GLE) setting has not yet been reported. The purpose of this work is to prove the validity of such a DFT for two solvable models: (i) a charged Brownian particle confined in a harmonic trap and driven by a prescribed time-dependent electric field, and (ii) an equivalent representation in which the center of the harmonic trap is arbitrarily dragged according to the same driving protocol. We take the magnetic field to be constant and directed along the $z$ axis, ${\bf B}=(0,0,B)$, and the electric field as ${\bf E}(t)={\bf E}_e(t)+{\bf E}_i(t)$, where ${\bf E}_e(t)$ is the external driving protocol and ${\bf E}_i(t)$ represents internal fluctuations (thermal noise). In the stochastic-thermodynamic sense, the reverse protocol is obtained by reversing the schedule of the externally controlled driving, ${\boldsymbol\lambda}_R(t)={\boldsymbol\lambda}_F(\tau-t)$. Here ${\boldsymbol\lambda}(t)\equiv{\bf E}_e(t)$ in the direct electric-field representation, with the analogous trap-center schedule used in the dragged-trap representation; $F$ and $R$ denote the forward and reverse protocols over the same time interval $\tau$, with ${\boldsymbol\lambda}_F(0)={\mathcal A}$ and ${\boldsymbol\lambda}_F(\tau)={\mathcal B}$. This protocol reversal is distinct from the microscopic time-reversal operation ${\mathcal T}({\bf r},{\bf v},t;{\bf B})=({\bf r},-{\bf v},-t;-{\bf B})$, under which velocities and magnetic fields are odd. Fluctuation relations for systems in a constant magnetic field can nevertheless be formulated without physically reversing the imposed magnetic field, provided the odd character of ${\bf B}$ under time reversal is treated consistently \cite{Coretti2020}. In the proof below, as in the Markovian charged-particle strategy of Ref.~\cite{Jimenez2010}, the constant magnetic field is kept as a fixed parameter of the GLE dynamics. The DFT is then obtained from the finite-time Gaussian position density entering the stochastic system entropy. In the present non-Markovian problem, this density is computed from the phase-space probability density (PSPD), which is obtained directly from the explicit solution of the GLE.

It will be shown that, starting from equilibrium canonical and Maxwellian initial distributions, the PSPD at time $t\ge 0$ remains Gaussian and factorizes into independent contributions. Moreover, in the absence of the time-dependent driving, the PSPD relaxes to a stationary distribution of the same form as the initial equilibrium one. This explicit time-dependent PSPD is the central ingredient that allows us to establish the DFT for the total entropy production in closed form.
 
This work is organized as follows. In Sec.~\ref{sec:2} we solve the GLE explicitly and obtain the PSPD. In Sec.~\ref{sec:3} we use this result to demonstrate the DFT for the total entropy production in two solvable models. Sec.~\ref{sec:illustrative-observables} presents representative observables and Monte Carlo simulations for an exponential memory kernel. Conclusions are presented in Sec.~\ref{sec:5}. The appendices contain the explicit solution, the Gaussian density reduction, and the covariance identities used in the present work.

\section{GLE in crossed electric and magnetic fields}
\label{sec:2}
\begin{figure*}[th]
\centering
\includegraphics[width=17cm, height=8cm]{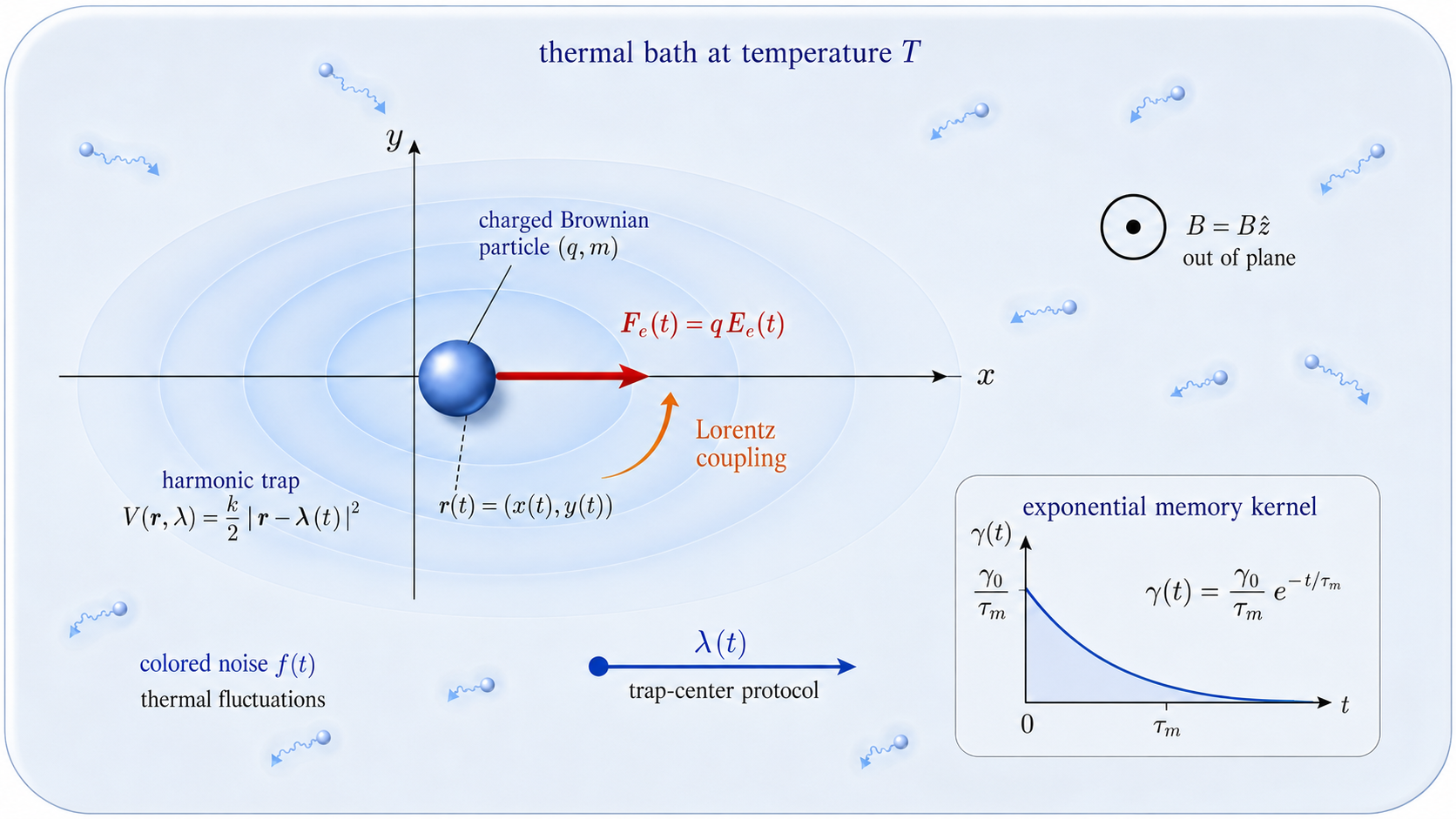}
\caption{Schematic of the model.  A charged Brownian particle in a harmonic trap is driven by the time-dependent electric force $\mathbf F_e(t)$ while a constant magnetic field $\mathbf B=B\hat{\mathbf z}$ couples the transverse motion.  The structured bath produces delayed friction through $\gamma(t)$ and colored fluctuations $\mathbf f(t)$.  The two driving protocols treated analytically are direct force driving and dragging of the trap center.}
\label{fig:model-schematic}
\end{figure*}

Figure~\ref{fig:model-schematic} summarizes the physical setting.  The particle is confined by a harmonic trap, driven by a prescribed electric force, and subjected to a constant magnetic field perpendicular to the plane of motion.  The bath is not assumed to respond instantaneously: its effect is represented by a memory kernel and a colored thermal force related by the fluctuation--dissipation theorem.

Consider a Brownian particle (an ion of mass $m$ and electric charge $q$) bound to a harmonic potential,
$V({\bf r})=\frac{1}{2}k |{\bf r}|^2$, of stiffness $k$, and under the influence of both a constant magnetic field pointing along the $z$-axis, namely ${\bf B}=(0,0,B)$,  and an external time-dependent electric field ${\bf E}_e(t)\equiv(E_x(t),E_y(t), E_z(t))$. In addition, fluctuations of the internal electric field ${\bf E}_i(t)$ (internal noise) account for the charged oscillator Brownian motion. If the tagged particle exchanges energy with the surrounding bath, its dynamics can be described by the following GLE:
\be
\ddot{\bf r}+\frac{1}{m}\int_0^t g(t-t')\,\dot{\bf r}(t')\, dt' +\frac{k}{m} {\bf r}-\frac{q}{m}\dot{\bf r} \times {\bf B}-\frac{q}{m}{\bf E}_e(t)=\frac{q}{m}{\bf E}_i(t)\,,  \label{gle} \ee
or
\be
\ddot{\bf r}+\int_0^t\gamma(t-t')\,\dot{\bf r}(t')\, dt' +\omega^2 {\bf r} -\frac{q}{m}\dot{\bf r} \times {\bf B}
-\frac{1}{m}{\bf F}_e(t)={\bf f}(t), 
\label{gle1} \ee
where ${\bf r}=(x,y,z)$ is the position vector, $\gamma(t)=g(t)/m$ is the friction memory kernel, and $\omega^2=k/m$ is the harmonic-oscillator frequency. The term ${\bf F}_e(t)\equiv q{\bf E}_e(t)$ denotes the external electric force, while ${\bf f}(t)=q{\bf E}_i(t)/m$ is the internal noise. In the absence of an external time-dependent force field, ${\bf f}(t)$ satisfies Kubo's second fluctuation--dissipation theorem \cite{Kubo1966},
\be
\langle f_i(t) f_j(t')\rangle={\frac{k_BT}{m}} \, \delta_{ij}\,\gamma(t-t') . \label{fdt1} \ee
This theorem guarantees that the stochastic process (\ref{gle1}) becomes stationary in the long-time
limit; this statement was proven explicitly in Ref.~\cite{Jimenez2015}.  Due to the orientation of the magnetic field, the GLE (\ref{gle1}) can be written in components as
\ba
\ddot x-\Omega \dot y+\omega^2 x +\int_0^t\gamma(t-t')\,\dot x(t')\, dt'
&-& {1\over m}F_{xe}(t) \cr\cr
&=&f_x(t) ,  \label{ddotx} \\
\ddot y+\Omega \dot x+\omega^2 y + \int_0^t\gamma(t-t')\,\dot y(t')\, dt' &-& {1\over m}F_{ye}(t) \cr\cr
&=&f_y(t) , \label{ddoty}  \\
\ddot z+\omega^2 z +\int_0^t\gamma(t-t')\,\dot z(t')\, dt' &-& {1\over m}F_{ze}(t) \cr\cr
&=&f_z(t)  ,  \label{ddotz}\ea
where $\Omega=qB/m$ is the cyclotron frequency. Along the $z$ axis, Eq.~(\ref{ddotz}) is independent of the magnetic field and decouples from the dynamics in the $(x,y)$ plane. We use the following notation throughout. Bold symbols without a tilde denote three-dimensional vectors, for example ${\bf r}=(x,y,z)$, ${\bf v}=(v_x,v_y,v_z)$, and ${\bf F}_e=(F_{xe},F_{ye},F_{ze})$. A tilde denotes projection onto the $xy$ plane, for example $\tilde{\bf r}=(x,y)$, $\tilde{\bf v}=(v_x,v_y)$, $\widetilde{\bf F}_e=(F_{xe},F_{ye})$, and $\tilde{\boldsymbol\lambda}=(\lambda_x,\lambda_y)$. Capital variables denote fluctuations about ensemble means:
\[
X=x-\langle x\rangle,\quad Y=y-\langle y\rangle,\quad Z=z-\langle z\rangle,\quad V_i=v_i-\langle v_i\rangle .
\]
Thus ${\bf R}=(X,Y,Z)$ and ${\bf S}=(V_x,V_y,V_z)$ denote the full fluctuation vectors, while $\widetilde{\bf R}=(X,Y)$ and $\widetilde{\bf S}=(V_x,V_y)$ denote their planar parts. We also write $\boldsymbol\Xi(t)=\langle \widetilde{\bf R}(t)\widetilde{\bf R}(t)\rangle$ for the planar positional covariance tensor.

The linear system (\ref{ddotx})--(\ref{ddotz}) is solved by Laplace transformation. The component solutions for the coordinates, velocities, means, and fluctuation variables are collected in \ref{app:explicit-solution}. For the derivation below, only two structural properties are needed: the solution is linear in the initial conditions, in the externally imposed force, and in the Gaussian colored noise; and the fluctuation variables satisfy an autonomous linear GLE independent of the driving protocol. These two facts imply that the phase-space distribution remains Gaussian for all finite times.

To prove the validity of the DFT for the total entropy production, we need the phase-space probability density (PSPD) associated with the GLE (\ref{gle1}). With the notation introduced above, the construction is straightforward. Since the solutions in Eqs.~(\ref{solx})--(\ref{solz}) and Eqs.~(\ref{solvx})--(\ref{solvz}) are linear functionals of the Gaussian colored noise ${\bf f}(t)$, the fluctuation vector $({\bf R},{\bf S})$ is Gaussian for each fixed time $t$. Therefore the PSPD $P({\bf r},{\bf v},t)\equiv P({\bf R},{\bf S})$ is a multivariate Gaussian distribution. Because the dynamics in the $xy$ plane decouples from the dynamics along the $z$ axis, this density factorizes as
\[
P({\bf R},{\bf S})=P_z(Z,V_z)\,P(\widetilde{\bf R},\widetilde{\bf S}) .
\]
The DFT proof below requires only the planar positional marginal $P(\tilde{\bf r},t)\equiv P(\widetilde{\bf R})$; the corresponding $z$-sector formulas are obtained by setting $\Omega=0$.

It is shown explicitly in \ref{app:pspd} that, for equilibrium initial conditions, the PSPD $P(\widetilde{\bf R},\widetilde{\bf S})$ reads
\be
P(\widetilde{\bf R},\widetilde{\bf S})={km\over (2\pi k_BT)^2}  {\rm exp}\bigg[-{(k|\widetilde{\bf R}|^2 +m|\widetilde{\bf S}|^2)\over 2 k_BT } \bigg] ,
  \label{gpd1} 
\ee

Here $\widetilde{\bf R}$ and $\widetilde{\bf S}$ are the planar fluctuation variables defined above. The corresponding marginal probability density $P(\tilde{\bf r},t)\equiv P(\widetilde{\bf R})$ is then
\be
P(\tilde{\bf r},t)={k\over 2\pi k_BT}\, \exp\left(- {k|\widetilde{\bf R}|^2\over 2k_BT } \right). \label{gpd2}
\ee

\section{DFT for the total entropy production} 
\label{sec:3}
We first set up the stochastic-thermodynamic quantities common to the two driving protocols. The discussion is restricted to the $xy$ plane, since the $z$ dynamics decouples from the magnetic field and does not affect the transverse entropy-production statistics. Within stochastic thermodynamics \cite{Seifert2005,Ohkuma2007,Sekimoto2010}, the heat $Q$ dissipated into the bath along a single stochastic trajectory $\tilde{\bf r}(t)$ over a finite time interval $\tau$ is related to the applied work $W$ and to the change in internal energy $\Delta U$ by the trajectory-level first-law balance
\be \label{fl}
Q=W-\Delta U .
\ee

For a protocol controlled by an externally prescribed planar parameter $\tilde{\boldsymbol\lambda}(t)$, the work $W\equiv W(\tau)$ is
\be
W=\int_0^{\tau} {\partial {\mathcal V}(\tilde{\bf r},\tilde{\boldsymbol\lambda}(t))\over\partial t} \, dt , \label{Wxy}
\ee
where ${\mathcal V}(\tilde{\bf r},\tilde{\boldsymbol\lambda}(t))$ is the effective external potential energy. We use this definition for two solvable protocols. In the direct-force protocol, the control parameter is the planar electric force, $\tilde{\boldsymbol\lambda}(t)=\widetilde{\bf F}_e(t)$. In the dragged-trap protocol, the control parameter is the trap center $\tilde{\boldsymbol\lambda}(t)$; equivalently, one may write $\widetilde{\bf F}_e(t)=k\tilde{\boldsymbol\lambda}(t)$ to express the trap displacement in force units.

The total entropy production is the sum of the medium entropy change, $\Delta s_m=Q/T$, and the system entropy change. The nonequilibrium Gibbs--Shannon entropy is
\be \label{S}
S(t)=- k_B\int P(\tilde{\bf r},t) \ln P(\tilde{\bf r},t) \, d\tilde{\bf r} 
=\langle s(t)\rangle ,
\ee
which defines the stochastic system entropy along a trajectory as
\be \label{s}
s(t)=- \ln P(\tilde{\bf r},t).
\ee
Here $P(\tilde{\bf r},t)$ is the marginal probability density of the particle position in the $xy$ plane, evaluated at the instantaneous trajectory point. Thus the change in system entropy between $0$ and $\tau$ is
\be \label{Ds1}
\Delta s=- \ln\left[ P(\tilde{\bf r},\tau)\over P(\tilde{\bf r}_0)\right] ,
\ee
and the total entropy production is
\be \label{Dst}
\Delta s_{tot}= \Delta s_m +\Delta s. 
\ee
Combining Eqs.~(\ref{fl})--(\ref{Ds1}) gives the trajectory functional
\be \label{Dst1}
\Delta s_{tot}={W-\Delta U\over T} 
-\ln\left[{P(\tilde{\bf r},\tau)\over P(\tilde{\bf r}_0)}\right]. 
\ee
This expression is the starting point for both protocols. It depends only on the work functional, the endpoint energy change, and the marginal densities at the beginning and at the end of the protocol.

The initial state is taken to be equilibrium, with the origin chosen so that $\langle \tilde{\bf r}_0\rangle=0$ and hence $\tilde{\bf r}_0=\widetilde{\bf R}_0=(X_0,Y_0)$. The initial marginal density is therefore
\be \label{Pr0}
P(\tilde{\bf r}_0)={k\over 2\pi k_BT}\, \exp\left(- {k |\tilde{\bf r}_0|^2\over 2k_BT} \right).  
\ee
Moreover, as shown in \ref{app:pspd}, the linear GLE dynamics preserves the Gaussian form of the fluctuation variables. At $t=\tau$, the marginal density can be written as
\be \label{Prtau}
P(\tilde{\bf r},\tau)={k\over 2\pi k_BT}\, \exp\left(- {k |\widetilde{\bf R}(\tau)|^2\over 2k_BT} \right).  
\ee
Equations~(\ref{Dst1}), (\ref{Pr0}), and (\ref{Prtau}) reduce the proof of the DFT to a Gaussian calculation: for each protocol we show that the total entropy production is a Gaussian random variable and that its variance is twice its mean.

\emph{Direct force protocol.} We first consider a Brownian particle in a harmonic trap driven by a prescribed planar electric force $\widetilde{\bf F}_e(t)=(F_{xe}(t),F_{ye}(t))$. In this protocol the externally controlled parameter is $\tilde{\boldsymbol\lambda}(t)=\widetilde{\bf F}_e(t)$, and the effective potential is
\[
{\mathcal V}(\tilde{\bf r}, \tilde{\boldsymbol\lambda}(t))={k\over2}|\tilde{\bf r}|^2-\widetilde{\bf F}_e(t)\cdot\tilde{\bf r}.
\]
Using Eq.~(\ref{Wxy}), the work performed over the interval $[0,\tau]$ is
\ba \label{Wxy1}
W&=&- \int_0^{\tau} \dot{\widetilde{\bf F}}_e(t)\cdot \tilde{\bf r}(t) \,dt
= - \int_0^{\tau} {\dot F}_{xe}(t)\,x(t)\, dt \cr\cr
&-&\int_0^{\tau} {\dot F}_{ye}(t)\,y(t)\, dt \, .
\ea
Assuming $\widetilde{\bf F}_e(0)=0$, the corresponding change in potential energy is
\ba \label{DU}
\Delta U &=& U(\tilde{\bf r}(\tau),\tau)-U(\tilde{\bf r}_0,0)\cr\cr
&=&  {1\over 2} k \left(|\tilde{\bf r}(\tau)|^2-|\tilde{\bf r}_0|^2\right)- \widetilde{\bf F}_e(\tau)\cdot\tilde{\bf r}(\tau),
\ea
and the mean work is
\ba \label{1mWt}
\langle W\rangle&=&- \int_0^{\tau} \dot {\widetilde{\bf F}}_e(t) \cdot \langle \tilde{\bf r}(t)\rangle\, dt\\
&=&-\int_0^{\tau} \bigg[\dot F_{xe}(t)\,\langle x(t)\rangle+\dot F_{ye}(t)\,\langle y(t)\rangle \bigg] dt \, .
\ea
The covariance calculation leading to the work variance is given in \ref{app:work-details}. For zero initial means, $\langle x_0\rangle=\langle y_0\rangle=\langle v_{x0}\rangle=\langle v_{y0}\rangle=0$, that calculation gives
\ba \label{varW5}
\sigma^2_W&=&{2\over\beta} \bigg[-\int_0^{\tau} \dot{\widetilde{\bf F}}_{e}(t)\cdot\langle \tilde{\bf r}(t)\rangle\, dt+ {1\over 2 m\omega^2}|\widetilde{\bf F}_e(\tau)|^2\bigg] \cr\cr
&=&{2\over\beta} \bigg[\langle W\rangle + {1\over 2 k}|\widetilde{\bf F}_e(\tau)|^2\bigg] . 
\ea

We now insert Eqs.~(\ref{Pr0}), (\ref{Prtau}), and (\ref{DU}) into Eq.~(\ref{Dst1}), so that the total entropy production for the direct-force protocol becomes
\ba 
\label{Dsta} 
\Delta s_{tot}&=&\frac{1}{T}\Bigg(W +\frac{1}{2}k|\langle\tilde {\bf r}(\tau)\rangle|^2
- k \tilde{\bf r}(\tau)\cdot \langle \tilde{\bf r}(\tau)\rangle\nonumber\\
&+&{\tilde{\bf r}}(\tau)\cdot \widetilde{\bf F}_e(\tau)  \Bigg).
\ea
Both $W$ and $\Delta s_{tot}$ are linear functionals of the Gaussian process $\tilde{\bf r}(t)$. Therefore $\Delta s_{tot}$ is Gaussian, with density
\ba
P(\Delta s_{tot})={1\over\sqrt{2\pi\, \sigma_s^2}} \exp\left[-{(\Delta s_{tot}-  \langle \Delta s_{tot}\rangle)^2 \over 2\sigma_s^2} \right],~~ \label{eppd}
\ea
where $\langle \Delta s_{tot}\rangle$ and $\sigma_s^2=\langle (\Delta s_{tot})^2\rangle- \langle \Delta s_{tot}\rangle^2$ are its mean and variance. From Eq.~(\ref{Dsta}),
\be \label{mDst}
\langle \Delta s_{tot}\rangle={1\over T} \left(\langle W\rangle - \frac{1}{2}k|\langle\tilde {\bf r}(\tau)\rangle|^2
+\langle \tilde{\bf r}(\tau)\rangle\cdot \widetilde{\bf F}_e(\tau)\right) ,
\ee
where $\langle W\rangle$ is given by Eq.~(\ref{1mWt}) and $\langle\tilde{\bf r}(\tau)\rangle$ follows from Eqs.~(\ref{solmx}) and (\ref{solmy}). \ref{app:work-details} gives the covariance identities required to evaluate the variance of Eq.~(\ref{Dsta}). Using those identities together with Eq.~(\ref{varW5}) yields
\ba
\sigma_s^2&=&{2}\left(\langle \Delta s_{tot}\rangle\right)\nonumber\\
&=&{2\over T}\left(\langle W\rangle - \frac{1}{2}k|\langle\tilde {\bf r}(\tau)\rangle|^2
+\langle \tilde{\bf r}(\tau)\rangle\cdot \widetilde{\bf F}_e(\tau)\right). ~~\label{Vep2}
\ea
Thus $\Delta s_{tot}$ is Gaussian and satisfies $\sigma_s^2=2\langle \Delta s_{tot}\rangle$. This mean--variance identity immediately gives the detailed fluctuation theorem
\begin{equation}
\frac{P(\Delta s_{tot})}{P(-\Delta s_{tot})}=e^{\Delta s_{tot}} ,  \label{dfts}
\end{equation}
for the non-Markovian dynamics governed by the GLE (\ref{gle1}) in the presence of a time-dependent electric force and a constant magnetic field. The integral fluctuation theorem follows by normalization:
\begin{equation}
\begin{split}
\langle e^{-\Delta s_{tot}}\rangle&=\int d\Delta s_{tot}\, e^{-\Delta s_{tot}} P(\Delta s_{tot})
\\
&=\int d\Delta s_{tot}\, P(-\Delta s_{tot})=1.    
\end{split}
\end{equation}

\emph{Dragged-trap protocol.} We next consider the second solvable protocol, in which the externally controlled parameter is the planar trap center $\tilde{\boldsymbol\lambda}(t)$. The harmonic potential is
\[
{\mathcal U}(\tilde{\bf r}, \tilde{\boldsymbol\lambda}(t))={k\over2}|\tilde{\bf r}-\tilde{\boldsymbol\lambda}(t)|^2 .
\]
For notational parallel with the direct-force protocol, we write
\[
\widetilde{\bf F}_e(t)=k\tilde{\boldsymbol\lambda}(t).
\]
This identification expresses the trap displacement in force units and allows the response formulas to be written in the same notation as above. For $\widetilde{\bf F}_e(0)=0$, the work is
\ba 
\hat W&=&\int_0^{\tau} {\partial {\mathcal U}(\tilde{\bf r},\tilde{\boldsymbol\lambda}(t))\over\partial t} \, dt \cr\cr
&=&- \int_0^{\tau} \dot{\widetilde{\bf F}}_e(t)\cdot \tilde{\bf r}(t) \,dt 
+ {|\widetilde{\bf F}_e(\tau)|^2\over 2k}, \label{hW1}
\ea
and the internal-energy change is
\be 
\Delta U = {k\over 2} \bigg|\tilde{\bf r}(\tau)-{\widetilde{\bf F}_e(\tau)\over k}\bigg|^2- {k\over 2}|\tilde{\bf r}_0|^2. \label{dDU}
\ee
The corresponding mean work is
\ba 
\langle \hat W\rangle
&=&- \int_0^{\tau} \dot{\widetilde{\bf F}}_e(t)\cdot \langle\tilde{\bf r}(t)\rangle \,dt 
+ {|\widetilde{\bf F}_e(\tau)|^2\over 2k}. \label{mhW}
\ea
Writing $\tilde{\bf r}(t)=\widetilde{\bf R}(t)+\langle \tilde{\bf r}(t)\rangle$, Eq.~(\ref{hW1}) becomes
\ba \label{hW2}
\hat W&=&- \int_0^{\tau} \dot {\widetilde{\bf F}}_e(t)\cdot \widetilde{\bf R}(t) \,dt
- \int_0^{\tau}\dot{\widetilde{\bf F}}_e(t)\cdot \langle\tilde{\bf r}(t)\rangle \,dt \cr\cr
&+& {|\widetilde{\bf F}_e(\tau)|^2\over 2k}
=- \int_0^{\tau} \dot {\widetilde{\bf F}}_e(t)\cdot \widetilde{\bf R}(t) \,dt+ \langle \hat W\rangle ,
\ea
so that
\be \label{VhW1}
\sigma^2_{\hat W}=\int_0^{\tau}\int_0^{\tau} \dot {\widetilde{\bf F}}_e(t)\cdot\langle \widetilde{\bf R}(t)\widetilde{\bf R}(t')\rangle\cdot \dot{\widetilde{\bf F}}_e(t') \,dt dt'.  
\ee
This is the same covariance functional that appears in the direct-force protocol, and the same calculation gives
\ba\label{VhW2}
\sigma^2_{\hat W}&=&{2\over\beta} \bigg[-\int_0^{\tau} \dot{\widetilde{\bf F}}_{e}(t)\cdot\langle \tilde{\bf r}(t)\rangle\, dt+ {1\over 2 k}|\widetilde{\bf F}_e(\tau)|^2\bigg] \cr\cr
&=&{2\over\beta} \langle \hat W\rangle .  
\ea

The entropy production for the dragged trap follows again from Eq.~(\ref{Dst1}), now using Eqs.~(\ref{hW1}) and (\ref{dDU}). One obtains
\be \label{Dhs1} 
\begin{split}
\Delta \hat s_{tot}&=\frac{1}{T}\Bigg(\hat W
+\frac{1}{2}k|\langle\tilde {\bf r}(\tau)\rangle|^2
- k \tilde{\bf r}(\tau)\cdot \langle \tilde{\bf r}(\tau)\rangle\\
&+{\tilde{\bf r}}(\tau)\cdot \widetilde{\bf F}_e(\tau)
- {1\over 2k}|\widetilde{\bf F}_e(\tau)|^2\Bigg) ,
\end{split}
\ee
with mean
\be \label{mDhs}
\begin{split}
\langle \Delta \hat s_{tot}\rangle&={1\over T} \Bigg(\langle \hat W\rangle
- \frac{1}{2}k|\langle\tilde {\bf r}(\tau)\rangle|^2\\
&+\langle \tilde{\bf r}(\tau)\rangle\cdot \widetilde{\bf F}_e(\tau)
- {1\over 2k}|\widetilde{\bf F}_e(\tau)|^2\Bigg) .
\end{split}
\ee
Because $\hat W$ and $\Delta \hat s_{tot}$ are again linear functionals of the Gaussian process $\tilde{\bf r}(t)$, the distribution $P(\Delta \hat s_{tot})$ has the Gaussian form in Eq.~(\ref{eppd}). Absorbing $k_B$ into $T$, its variance can be written as
\ba
\sigma_{\hat s}^2&=&{1\over T^2}\bigg(
[\langle \hat W \tilde{\bf r}(\tau)\rangle-\langle \hat W \rangle\langle\tilde{\bf r}(\tau)\rangle ]\cdot[2\widetilde{\bf F}_e(\tau)- 2k\langle\tilde{\bf r}(\tau)\rangle] \cr\cr
&+& \widetilde{\bf F}_e(\tau)\cdot {\boldsymbol\Xi}(\tau)\cdot \widetilde{\bf F}_e(\tau)+\sigma^2_{\hat W} \nonumber\\
&-&2k\,\langle \tilde{\bf r}(\tau)\rangle \cdot {\boldsymbol\Xi}(\tau)\cdot \widetilde{\bf F}_e(\tau)
+k^2\langle \tilde{\bf r}(\tau)\rangle\cdot {\boldsymbol\Xi}(\tau)\cdot \langle \tilde{\bf r}(\tau)\rangle  \bigg). \nonumber\\ 
\label{Vhs1}
\ea
The mixed covariance in Eq.~(\ref{Vhs1}) reduces to
\ba \label{hWR1}
\langle \hat W \widetilde{\bf R}(\tau)\rangle
&=&-\left\langle\left( \int_0^{\tau} \dot {\widetilde{\bf F}}_e(t) \cdot \tilde{\bf r}(t)\, dt
+{|\widetilde{\bf F}_e(\tau)|^2\over 2k}\right)\widetilde{\bf R}(\tau)\right\rangle\cr\cr
&=& -\int_0^{\tau} \dot {\widetilde{\bf F}}_e(t) \cdot \langle \widetilde{\bf R}(\tau)\widetilde{\bf R}(t)\rangle\, dt ,
\ea
where $\langle \widetilde{\bf R}(\tau)\rangle={\bf 0}$. This is the same covariance obtained in the direct-force calculation, and hence
\be \label{hWR2}
\langle \hat W \widetilde{\bf R}(\tau)\rangle=-{T\over k} \big[\widetilde{\bf F}_{e}(\tau)-k\langle\tilde{\bf r}(\tau)\rangle \big].  
\ee
Substitution of Eqs.~(\ref{VhW2}) and (\ref{hWR2}), together with the covariance tensor identities for ${\boldsymbol\Xi}(\tau)$, gives
\ba
\sigma_{\hat s}^2&=&{2\over T}\Bigg(\langle \hat W\rangle
- \frac{1}{2}k|\langle\tilde {\bf r}(\tau)\rangle|^2+\langle \tilde{\bf r}(\tau)\rangle\cdot \widetilde{\bf F}_e(\tau)
- {|\widetilde{\bf F}_e(\tau)|^2\over 2k}\Bigg)\cr\cr
&=&2\langle \Delta \hat s_{tot}\rangle . \label{Vhs2}
\ea
Thus the dragged-trap entropy production is also Gaussian and satisfies $\sigma_{\hat s}^2=2\langle \Delta \hat s_{tot}\rangle$. Consequently the same DFT and IFT hold for $\Delta \hat s_{tot}$. The two protocols therefore share the same structural mechanism: linear GLE dynamics preserves Gaussianity, and the covariance identities enforce the variance--mean relation required by the entropy-production fluctuation theorem.

\section{Representative observables and Monte Carlo simulations}
\label{sec:illustrative-observables}
The analytical derivation above is independent of the particular functional form of the memory kernel, provided the GLE remains linear and the fluctuation--dissipation relation is satisfied. To make the physical content more transparent, we show representative observables for an exponentially decaying memory kernel,
\be
\gamma(t)={\gamma_0\over \tau_m}e^{-t/\tau_m}.
\label{exponential-memory}
\ee
This choice is not required by the theorem; it is used to visualize how memory and the magnetic field affect the driven response and the entropy-production symmetry. The magnetic field produces a transverse response, so that a drive along $x$ generates motion in the $y$ direction, while the memory time changes the relaxation lag and the shape of the trajectory in the plane.

For the exponential kernel in Eq.~(\ref{exponential-memory}), the non-Markovian convolution can be represented by a Markovian embedding with auxiliary bath variables $h_x(t)$ and $h_y(t)$. In the direct-force representation, the embedded dynamics is
\begin{eqnarray}
\dot{x}&=&v_x,\qquad \dot{y}=v_y,\nonumber\\
\dot{v}_x&=&-kx/m+\Omega v_y-h_x+F_{xe}(t)/m,\nonumber\\
\dot{v}_y&=&-ky/m-\Omega v_x-h_y+F_{ye}(t)/m,\nonumber\\
\dot{h}_i&=&-h_i/\tau_m+(\gamma_0/\tau_m)v_i
+{\sqrt{2\gamma_0 k_BT}\over \tau_m}\eta_i(t),\quad i=x,y ,
\label{embedded-gel-numerics}
\end{eqnarray}
where $\eta_x(t)$ and $\eta_y(t)$ are independent Gaussian white noises satisfying $\langle\eta_i(t)\eta_j(t')\rangle=\delta_{ij}\delta(t-t')$. The initial ensemble is sampled from the equilibrium distribution of the embedded linear system. This preparation fixes both the particle variables and the auxiliary bath variables consistently with the fluctuation--dissipation relation. For the dragged-trap protocol, the same embedded linear equations are used with the force-unit notation $\widetilde{\bf F}_e(t)=k\tilde{\boldsymbol\lambda}(t)$. This keeps the deterministic mean equations in the same form, while the thermodynamic work remains the dragged-trap work defined by the time dependence of the harmonic center. In both representations the protocol shifts the mean trajectory, while the fluctuation covariance around the mean is determined by the autonomous linear fluctuation dynamics.

\begin{figure}[H]
\centering
\includegraphics[width=0.98\columnwidth]{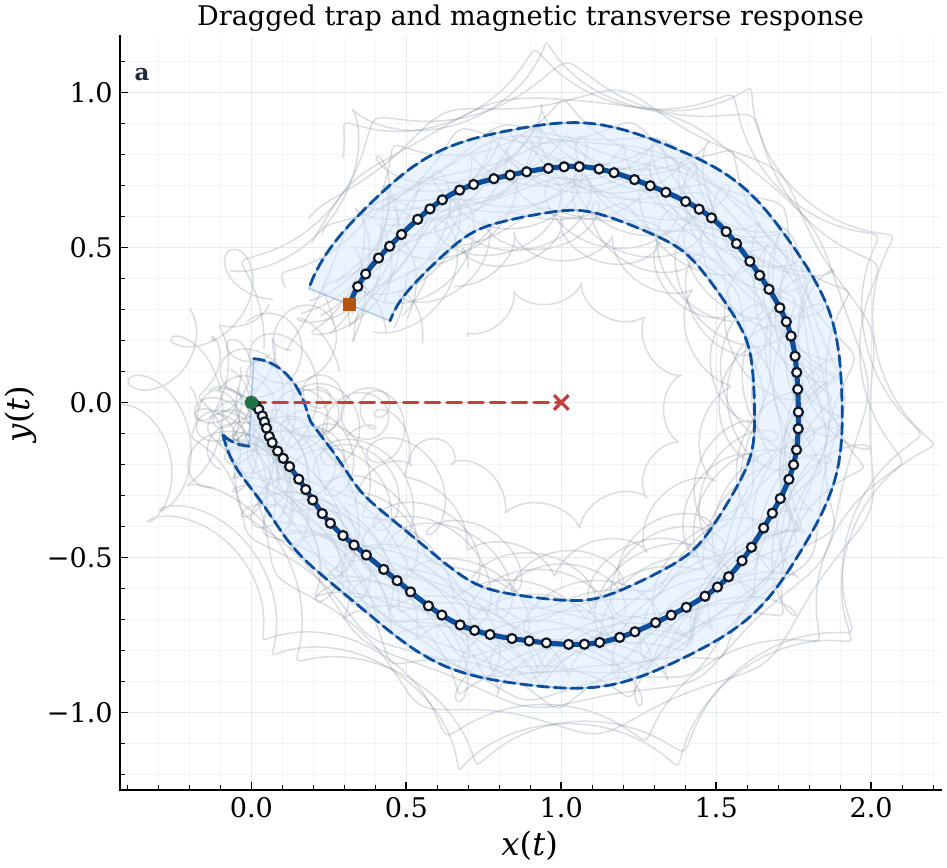}
\caption{Trajectory-space response for the exponential kernel in Eq.~(\ref{exponential-memory}). Thin gray curves (\protect\raisebox{0.35ex}{\textcolor{gray}{\rule{1.2em}{0.4pt}}}) show representative stochastic trajectories in the $xy$ plane; the light-blue shaded tube (\protect\raisebox{0.1ex}{\textcolor{blue!18}{\rule{1.1em}{0.65em}}}) is the one-standard-deviation trajectory spread estimated from the Monte Carlo covariance normal to the mean path; the dashed blue curves (\protect\raisebox{0.35ex}{\textcolor{blue}{\rule{0.45em}{0.65pt}\hspace{0.15em}\rule{0.45em}{0.65pt}}}) show the corresponding theoretical boundaries; the dashed red line (\protect\raisebox{0.35ex}{\textcolor{red}{\rule{0.45em}{0.65pt}\hspace{0.15em}\rule{0.45em}{0.65pt}}}) is the dragged trap-center path; the solid blue curve (\protect\raisebox{0.35ex}{\textcolor{blue}{\rule{1.2em}{0.8pt}}}) is the deterministic mean trajectory; open black circles ($\circ$) are Monte Carlo estimates of the mean at selected times; the green filled circle (\textcolor{green!50!black}{$\bullet$}) marks the initial mean position; the orange square (\textcolor{orange}{$\blacksquare$}) marks the final mean position; and the red cross (\textcolor{red}{$\times$}) marks the final trap center, $\lambda_f=F_0/k=1$. The parameters are $m=1$, $k=2$, $k_BT=0.04$, $\gamma_0=0.05$, $\tau_m=0.8$, $\Omega=4$, $F_0=2$, $\tau_d=4$, $\tau=15$, $\Delta t=5\times10^{-3}$, and $3\times10^4$ trajectories; $36$ individual trajectories are displayed for visual clarity.}
\label{fig:mean-response-mc}
\end{figure}

The two numerical illustrations below have different purposes and therefore use different representative parameter sets. Figure~\ref{fig:mean-response-mc} is designed to display the geometry of the stochastic motion in the $xy$ plane. For that purpose we use the dragged-trap protocol discussed in Sec.~\ref{sec:3}, specialized to a constant-velocity displacement of the trap center followed by a hold stage:
\begin{equation}
\lambda_x(t)=
\begin{cases}
\lambda_f t/\tau_d, & 0\leq t\leq \tau_d,\\
\lambda_f, & \tau_d<t\leq \tau,
\end{cases}
\qquad
\lambda_y(t)=0,
\qquad
\lambda_f={F_0\over k}.
\end{equation}
Equivalently, the longitudinal force associated with the displaced harmonic center is $F_{xe}(t)=k\lambda_x(t)$, with $F_{ye}(t)=0$. The trap is therefore dragged from $\lambda_x=0$ to $\lambda_x=\lambda_f$ during $0\leq t\leq \tau_d$ and then held fixed until the final observation time. We use $m=1$, $k=2$, $k_BT=0.04$, $\gamma_0=0.05$, $\tau_m=0.8$, $\Omega=4$, $F_0=2$, so that $\lambda_f=1$, with dragging time $\tau_d=4$, final observation time $\tau=15$, time step $\Delta t=5\times10^{-3}$, and $3\times10^4$ trajectories. These parameters place the response in an underdamped, strongly magnetized regime in which the particle lags behind the moving trap center and the Lorentz-induced curvature of the mean trajectory is visible.

In Fig.~\ref{fig:mean-response-mc}, the shaded tube is obtained directly from the Monte Carlo covariance of the trajectory ensemble. More precisely, in the planar notation introduced above, let ${\boldsymbol\mu}_{\rm MC}(t)=(\langle x(t)\rangle_{\rm MC},\langle y(t)\rangle_{\rm MC})$ be the Monte Carlo mean trajectory and let $\Sigma_{\rm MC}(t)$ be the sample covariance matrix of the planar position $(x(t),y(t))$. The displayed width is the one-standard-deviation projection
\begin{equation}
\sigma_{\perp,{\rm MC}}(t)=
\left[{\bf n}_{\rm MC}(t)^{T}\Sigma_{\rm MC}(t){\bf n}_{\rm MC}(t)\right]^{1/2},
\end{equation}
where ${\bf n}_{\rm MC}(t)$ is the unit normal to the Monte Carlo mean trajectory. The dashed boundary curves are obtained from the corresponding theoretical covariance of the embedded linear system, evaluated along the deterministic mean trajectory. Thus the shaded region represents the physical trajectory-to-trajectory spread of the Brownian ensemble.

Figure~\ref{fig:mean-response-mc} illustrates the physical mechanism that enters the analytical response functions. As the harmonic center is dragged along $x$, the Brownian particle does not simply follow it longitudinally: memory produces a lag relative to the imposed motion, while the Lorentz coupling rotates the response and generates a transverse displacement. The individual trajectories fluctuate around the mean path with a finite thermal width, and the Monte Carlo tube agrees with the theoretical covariance boundaries. The smoothness of the displayed position trajectories is consistent with the underdamped exponential-memory dynamics: the noise acts through the auxiliary bath variables and is filtered by the inertial motion before appearing in the particle position.

\begin{figure*}[t!]
\centering
\includegraphics[width=17cm, height=6cm]{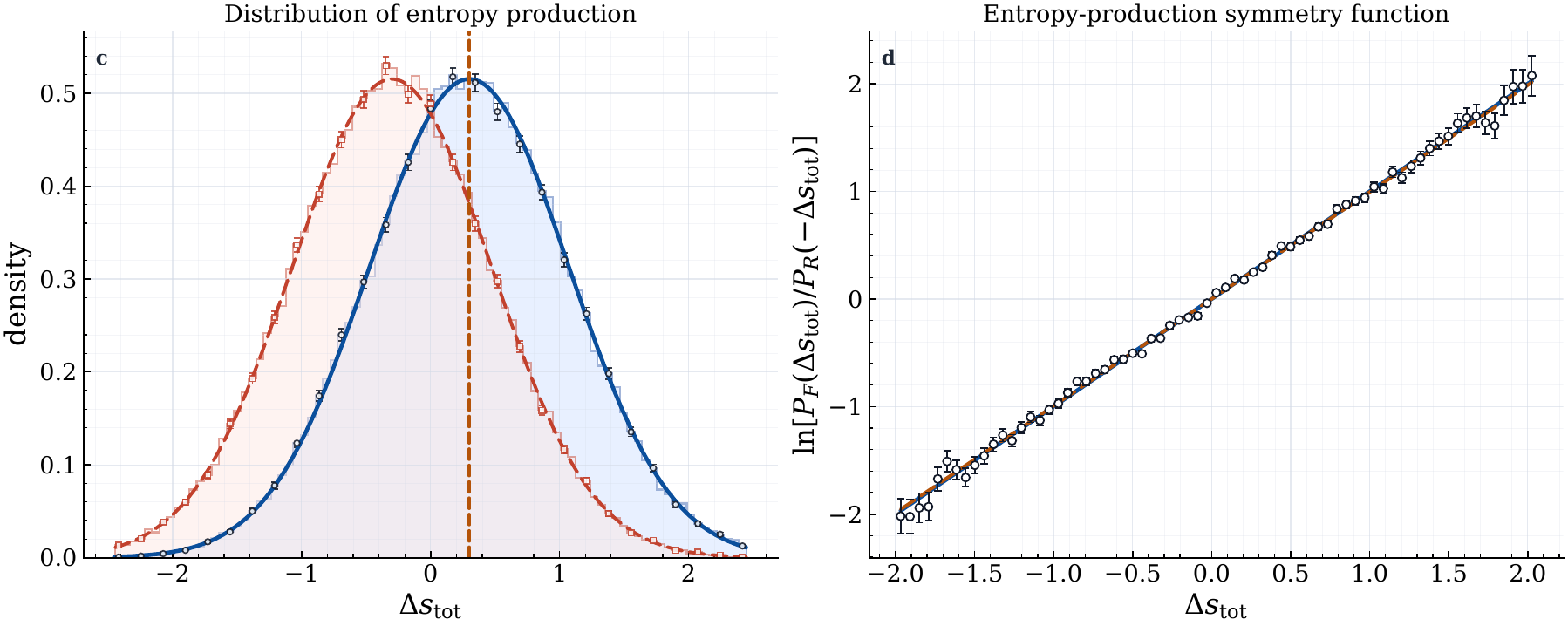}
\caption{Entropy-production distribution and symmetry function for the direct linear force protocol. Left panel shows the forward Monte Carlo histogram of $\Delta s_{\rm tot}$ (\protect\raisebox{0.1ex}{\textcolor{blue!18}{\rule{1.1em}{0.65em}}}) and the reverse Monte Carlos histogram  (\protect\raisebox{0.1ex}{\textcolor{red!18}{\rule{1.1em}{0.65em}}}). The corresponding analytical Gaussian curves are the forward Gaussian (\protect\raisebox{0.35ex}{\textcolor{blue}{\rule{1.2em}{0.8pt}}}) and the reverse Gaussian (\protect\raisebox{0.35ex}{\textcolor{red}{\rule{0.45em}{0.65pt}\hspace{0.15em}\rule{0.45em}{0.65pt}}}); the vertical dashed orange line (\protect\raisebox{0.35ex}{\textcolor{yellow!35!red!70!black}{\rule{0.45em}{0.65pt}\hspace{0.15em}\rule{0.45em}{0.65pt}}}) marks $\langle\Delta s_{\rm tot}\rangle_F$. Open black circles ($\circ$) are forward-bin estimates with errors, and open red squares (\textcolor{red}{$\square$}) are reverse-bin estimates with errors. Right Panel shows the empirical symmetry function $\ln[P_F(\Delta s_{\rm tot})/P_R(-\Delta s_{\rm tot})]$: open black circles ($\circ$) are the Monte Carlo histogram estimates, the solid blue line (\protect\raisebox{0.35ex}{\textcolor{blue}{\rule{1.2em}{0.8pt}}}) is the DFT prediction with unit slope, and the dashed orange line (\protect\raisebox{0.35ex}{\textcolor{yellow!35!red!70!black}{\rule{0.45em}{0.65pt}\hspace{0.15em}\rule{0.45em}{0.65pt}}}) is a linear fit. The parameters are $m=1$, $k=1$, $k_BT=1$, $\gamma_0=1.2$, $\tau_m=0.9$, $\Omega=1$, $F_0=1.4$, $\tau=6$, $\Delta t=10^{-2}$, and $10^5$ trajectories for each protocol.}
\label{fig:entropy-symmetry-mc}
\end{figure*}

Fig.~\ref{fig:entropy-symmetry-mc} shows the entropy-production result for the direct-force protocol of Sec.~\ref{sec:3}, rather than the dragged-trap protocol used for the trajectory-space visualization. For the forward experiment we take
\begin{equation}
F_{xe}^{F}(t)=F_0 t/\tau,\qquad F_{ye}^{F}(t)=0,
\end{equation}
and for the reverse experiment we take
\begin{equation}
F_{xe}^{R}(t)=F_0(1-t/\tau),\qquad F_{ye}^{R}(t)=0.
\end{equation}
The magnetic-field orientation is kept fixed in this numerical forward--reverse comparison. The representative dimensionless parameters used for Fig.~\ref{fig:entropy-symmetry-mc} are $m=1$, $k=1$, $k_BT=1$, $\gamma_0=1.2$, $\tau_m=0.9$, $\Omega=1$, $F_0=1.4$, $\tau=6$, $\Delta t=10^{-2}$, and $10^5$ trajectories for each of the forward and reverse ensembles.

The entropy-production statistics in Fig.~\ref{fig:entropy-symmetry-mc} are consistent with the Gaussian structure described in the previous sections. The forward and reverse distributions have the Gaussian form predicted by Eq.~(\ref{eppd}), with mean and variance constrained by Eqs.~(\ref{mDst}) and (\ref{Vep2}). The symmetry function is linear over the range where both the forward and reverse histograms are sampled with sufficient statistics, and its slope is close to the unit value required by the DFT in Eq.~(\ref{dfts}). Thus the numerical results illustrate two complementary aspects of the theory: memory and magnetic coupling shape the observable trajectory response, while the total entropy production retains the fluctuation-theorem symmetry imposed by the Gaussian structure of the linear GLE.

\section{Concluding Remarks}
\label{sec:5}
In this work we established, within a fully non-Markovian framework, the detailed fluctuation theorem  for the total entropy production of a charged Brownian particle in a harmonic trap driven by a time-dependent electric field and subjected to a constant transverse magnetic field. Our approach is based directly on the generalized Langevin equation and exploits the fact that the dynamics is linear and the noise is Gaussian, so that the phase-space probability density remains Gaussian for all times. For equilibrium initial conditions---specifically, an initial distribution given by the product of a canonical distribution in position and a Maxwellian in velocity---we obtained the explicit marginal density $P(\tilde{\bf r},t)$ needed to define the trajectory-dependent system entropy, and showed that it retains a canonical form when written in terms of the fluctuation variables about the mean trajectory. This structure is similar in spirit, though not identical in detail, to the Markovian result reported in Ref.~\cite{Jimenez2010}.

We considered two solvable driving protocols: (i) a harmonically trapped particle under a time-dependent force field, and (ii) the dragging of the trap center with an arbitrary time dependence. In both cases, the total entropy production is a Gaussian random variable and satisfies the relation between variance and mean that implies the DFT, and consequently the corresponding integral fluctuation theorem. An interesting direction for future work is to confront these predictions with experiments in which non-Markovian effects are relevant (e.g.\ viscoelastic environments) and where magnetic fields play an essential role; to our knowledge, systematic experimental tests of fluctuation theorems in such magnetically driven non-Markovian settings are still scarce. The simulations in Sec.~\ref{sec:illustrative-observables} illustrate this reconstruction procedure in a controlled setting: the driven response is sampled, work and entropy production are reconstructed from the trajectory ensemble, and the empirical symmetry function is compared with the DFT prediction.

\section*{Acknowledgments}
L.C.G.M. acknowledges a doctoral fellowship from SECIHTI through the Becas Nacionales para Estudios de Posgrado program for doctoral studies at UAM-Iztapalapa. I.P.C. and J.I.J.A. acknowledge support from SNII-SECIHTI. I.P.C. also acknowledges financial support from SECIHTI through research grant CBF-2025-I-3911.

\bibliographystyle{elsarticle-num}
\bibliography{references}

\newpage

\appendix
\section{Explicit linear solution and fluctuation variables}
\label{app:explicit-solution}
This appendix collects the component solution of the GLE, the corresponding mean values, and the fluctuation variables used in the main text. The explicit formulas are retained here to make the derivation reproducible, while Sec.~\ref{sec:3} uses only their structural consequences: linearity in the initial conditions, linearity in the external protocol, and linearity in the Gaussian colored noise.

The solution of Eqs.~(\ref{ddotx})--(\ref{ddotz}) can be obtained by a Laplace transform, which yields
\ba
&&x(t)=x_0[\chi_0(t)+\Omega^2\omega^2 \chi_2(t)] - y_0 \Omega\omega^2 H_1(t)\cr\cr
&+&v_{x0} [H_0(t)-\Omega^2{\mathcal H}_2(t)] +v_{y0}\Omega {\mathcal H}_1(t) \cr\cr
 &+&{1\over m}\int_0^t H_0(t-t') F_{xe}(t') dt' \cr\cr
 &-& {\Omega^2\over m} \int_0^t {\mathcal H}_2(t-t')  F_{xe}(t') dt' \cr\cr
&+& {\Omega\over m} \int_0^t {\mathcal H}_1(t-t')  F_{ye}(t') dt' +\int_0^t H_0(t-t') f_x(t') dt' \cr\cr
&-&\Omega^2 \int_0^t {\mathcal H}_2(t-t')  f_x(t') dt' +\Omega \int_0^t {\mathcal H}_1(t-t')  f_y(t') dt',\nonumber\\
 \label{solx}  
\ea
\ba
&& y(t)=y_0[\chi_0(t)+\Omega^2\omega^2 \chi_2(t)] + x_0 \Omega\omega^2 H_1(t)\cr\cr
&+&v_{y0} [H_0(t)-\Omega^2{\mathcal H}_2(t)] -v_{x0}\Omega {\mathcal H}_1(t) \cr\cr
&+& {1\over m}\int_0^t H_0(t-t') F_{ye}(t') dt' \cr\cr
&-&{\Omega^2\over m} \int_0^t {\mathcal H}_2(t-t')  F_{ye}(t') dt' \cr\cr
&-& {\Omega\over m} \int_0^t {\mathcal H}_1(t-t')  F_{xe}(t') dt' + \int_0^t H_0(t-t') f_y(t') dt'\cr\cr
&-&\Omega^2 \int_0^t {\mathcal H}_2(t-t')  f_y(t') dt' -\Omega \int_0^t {\mathcal H}_1(t-t')  f_x(t') dt', \nonumber\\
  \label{soly}  \ea
\ba
z(t)&=&z_0\chi_0(t)+v_{z0} H_0(t) +{1\over m}\int_0^t H_0(t-t') F_{ze}(t') dt' \cr\cr
&+&\int_0^t H_0(t-t') f_z(t') dt'.
  \label{solz} 
\ea
The mean values are obtained by averaging Eqs.~(\ref{solx})--(\ref{solz}) and using $\langle {\bf f}(t)\rangle={\bf 0}$:
\ba
&&\langle x(t)\rangle=\langle x_0\rangle [\chi_0(t)+\Omega^2\omega^2 \chi_2(t)] - \langle y_0\rangle \Omega\omega^2 H_1(t)\cr\cr
&+&\langle v_{x0}\rangle [H_0(t)-\Omega^2{\mathcal H}_2(t)] +\langle v_{y0}\rangle \Omega {\mathcal H}_1(t) \cr\cr
 &+&{1\over m}\int_0^t H_0(t-t') F_{xe}(t') dt' \cr\cr
 &-& {\Omega^2\over m} \int_0^t {\mathcal H}_2(t-t')  F_{xe}(t') dt'\cr\cr
 &+& {\Omega\over m} \int_0^t {\mathcal H}_1(t-t')  F_{ye}(t') dt',    \label{solmx}  \ea
\ba
&&\langle y(t)\rangle =\langle y_0\rangle[\chi_0(t)+\Omega^2\omega^2 \chi_2(t)] + \langle x_0\rangle \Omega\omega^2 H_1(t)\cr\cr
&+&\langle v_{y0}\rangle [H_0(t)-\Omega^2{\mathcal H}_2(t)] -\langle v_{x0}\rangle \Omega {\mathcal H}_1(t) \cr\cr
&+& {1\over m}\int_0^t H_0(t-t') F_{ye}(t') dt' \cr\cr
&-&{\Omega^2\over m} \int_0^t {\mathcal H}_2(t-t')  F_{ye}(t') dt' \cr\cr
&-& {\Omega\over m} \int_0^t {\mathcal H}_1(t-t')  F_{xe}(t') dt' ,  \label{solmy}  \ea
\ba
\langle z(t)\rangle&=&\langle z_0\rangle \chi_0(t)+\langle v_{z0}\rangle H_0(t) \cr\cr
&+&{1\over m}\int_0^t H_0(t-t') F_{ze}(t') dt' .  \label{solmz} \ea
Here $x_0=x(0)$, $y_0=y(0)$, and $z_0=z(0)$ denote the initial particle position, while
$v_{x0}=v_x(0)$, $v_{y0}=v_y(0)$, and $v_{z0}=v_z(0)$ are the initial velocities. The auxiliary functions
$\chi_0(t)$ and $\chi_2(t)$ are defined as
\be
\chi_0(t)=1-\omega^2\int_0^t H_0(t') dt'    \label{chi0}\, ,
\ee
\be 
\chi_2(t)=\int_0^t {\mathcal H}_2(t') dt'  \label{chi2}\, .
\ee 
The functions $H_0(t)$, ${\mathcal H}_1(t)$, and ${\mathcal H}_2(t)$ are the inverse Laplace transforms of $\hat H_0(s)$, $\hat{\mathcal H}_1(s)$, and $\hat{\mathcal H}_2(s)$, respectively, i.e., $H_0(t)={\mathcal L}^{-1}\{\hat H_0(s)\}$, ${\mathcal H}_1(t)={\mathcal L}^{-1}\{\hat{\mathcal H}_1(s)\}$, and
${\mathcal H}_2(t)={\mathcal L}^{-1}\{\hat{\mathcal H}_2(s)\}$. We define $\hat{\mathcal H}_1(s)=s\hat H_1(s)$ and $\hat{\mathcal H}_2(s)=s^2 \hat H_2(s)$, with
\be
\hat H_0(s)={1\over s^2+s\hat\gamma(s)+\omega^2}   \label{hatH0}\, , \ee
\be
\hat H_1(s)={1\over (s^2+s\hat\gamma(s)+\omega^2)^2+(\Omega s)^2}   \label{hatH1}\, , \ee
\be
\hat H_2(s)={1\over (s^2+s\hat\gamma(s)+\omega^2)
[(s^2+s\hat\gamma(s)+\omega^2)^2+(\Omega s)^2]}   \label{hatH2}\, ,\ee
where $\hat\gamma(s)$ is the Laplace transform of $\gamma(t)$. Differentiating the coordinate solution gives the velocity ${\bf v}=(v_x,v_y,v_z)$:
\ba
&&v_x(t)=x_0[-\omega^2H_0(t)+\Omega^2\omega^2 {\mathcal H}_2(t)] 
- y_0 \Omega\omega^2 {\dot H}_1(t) \cr\cr
&+&v_{x0} [{\dot H}_0(t)-\Omega^2 \dot{\mathcal H}_2(t)] +v_{y0}\Omega \dot{\mathcal H}_1(t) \cr\cr
&+& {1\over m}\int_0^t  {\dot H}_0(t-t') F_{xe}(t') dt' \cr\cr
&-&{\Omega^2\over m} \int_0^t \dot{\mathcal H}_2(t-t')  F_{xe}(t') dt'  \cr\cr
&+& {\Omega\over m} \int_0^t \dot{\mathcal H}_1(t-t')  F_{ye}(t') dt' +\int_0^t {\dot H}_0(t-t') f_x(t') dt' \cr\cr
&-&\Omega^2 \int_0^t \dot{\mathcal H}_2(t-t')  f_x(t') dt' + \Omega \int_0^t \dot{\mathcal H}_1(t-t')  f_y(t') dt',  \nonumber\\ \label{solvx} \ea
\ba
&&v_y(t)=y_0[-\omega^2H_0(t)+\Omega^2\omega^2 {\mathcal H}_2(t)] 
+ x_0 \Omega\omega^2 {\dot H}_1(t) \cr\cr
&+&v_{y0} [{\dot H}_0(t)-\Omega^2 \dot{\mathcal H}_2(t)] -v_{x0}\Omega \dot{\mathcal H}_1(t) \cr\cr
&+& {1\over m}\int_0^t {\dot H}_0(t-t') F_{ye}(t') dt' \cr\cr
&-&{\Omega^2\over m} \int_0^t \dot{\mathcal H}_2(t-t')  F_{ye}(t') dt' \cr\cr
&-& {\Omega\over m} \int_0^t \dot{\mathcal H}_1(t-t')  F_{xe}(t') dt' + \int_0^t {\dot H}_0(t-t') f_y(t') dt' \cr\cr
&-& \Omega^2 \int_0^t \dot{\mathcal H}_2(t-t')  f_y(t') dt'
- \Omega \int_0^t \dot{\mathcal H}_1(t-t') f_x(t') dt'  ,  \nonumber\\ \label{solvy}  \ea
\ba
&&v_z(t)=-\omega^2 z_0H_0(t) +v_{z0} {\dot H}_0(t) \cr\cr
&+&{1\over m} \int_0^t {\dot H}_0(t-t') F_{ze}(t') dt' + \int_0^t {\dot H}_0(t-t') f_z(t') dt', \nonumber \\
\label{solvz} 
\ea
The corresponding mean values are 
\ba
&&\langle v_x(t)\rangle=\langle x_0\rangle[-\omega^2H_0(t)+\Omega^2\omega^2 {\mathcal H}_2(t)] \cr\cr
&-& \langle y_0\rangle \Omega\omega^2 {\dot H}_1(t) + \langle v_{x0}\rangle [{\dot H}_0(t)-\Omega^2 \dot{\mathcal H}_2(t)] \cr\cr
&+&\langle v_{y0}\rangle \Omega \dot{\mathcal H}_1(t)
+ {1\over m}\int_0^t {\dot H}_0(t-t') F_{xe}(t') dt' \cr\cr
&-&{\Omega^2\over m} \int_0^t \dot{\mathcal H}_2(t-t')  F_{xe}(t') dt' \cr\cr
&+& {\Omega\over m} \int_0^t \dot{\mathcal H}_1(t-t')  F_{ye}(t') dt' ,  \label{solmvx} 
\ea
\ba
&&\langle v_y(t)\rangle=\langle y_0\rangle[-\omega^2H_0(t)+\Omega^2\omega^2 {\mathcal H}_2(t)] \cr\cr
&+& \langle x_0\rangle \Omega\omega^2 {\dot H}_1(t) 
+\langle v_{y0}\rangle [{\dot H}_0(t)-\Omega^2 \dot{\mathcal H}_2(t)] \cr\cr
&-&\langle v_{x0}\rangle\Omega \dot{\mathcal H}_1(t) 
+ {1\over m}\int_0^t {\dot H}_0(t-t') F_{ye}(t') dt' \cr\cr
&-&{\Omega^2\over m} \int_0^t \dot{\mathcal H}_2(t-t')  F_{ye}(t') dt' \cr\cr
&-& {\Omega\over m} \int_0^t \dot{\mathcal H}_1(t-t')  F_{xe}(t') dt' ,\label{solmvy}  
\ea
\ba
\langle v_z(t)\rangle&=&-\langle z_0\rangle\omega^2 H_0(t)+\langle v_{z0}\rangle {\dot H}_0(t) \cr\cr 
&+& {1\over m}\int_0^t {\dot H}_0(t-t') F_{ze}(t') dt'.  \label{solmvz} 
\ea

The other appropriate variables that we will use in this work are the fluctuations $X=x-\langle x\rangle$,
$Y=y-\langle y\rangle$, and $Z=z-\langle z\rangle$, which satisfy
\ba
&&\ddot X-\Omega \dot Y+\omega^2 X +\int_0^t\gamma(t-t')\,\dot X(t')\, dt' =f_x(t) , ~~~ \label{ddotX}  \\ 
&&\ddot Y +\Omega \dot X+\omega^2 Y + \int_0^t\gamma(t-t')\,\dot Y(t')\, dt'=f_y(t) , ~~~ \label{ddotY}  \\
&&\ddot Z+\omega^2 Z +\int_0^t\gamma(t-t')\,\dot Z(t')\, dt' =f_z(t)  .  \label{ddotZ}\ea
The corresponding solutions are
\ba
&&X(t)=X_0[\chi_0(t)+\Omega^2\omega^2 \chi_2(t)] - Y_0 \Omega\omega^2 H_1(t)\cr\cr
&+&V_{x0} [H_0(t)-\Omega^2{\mathcal H}_2(t)] +V_{y0}\Omega {\mathcal H}_1(t) \cr\cr
 &+&\int_0^t H_0(t-t') f_x(t') dt' 
-\Omega^2 \int_0^t {\mathcal H}_2(t-t')  f_x(t') dt' \cr\cr
&+&\Omega \int_0^t {\mathcal H}_1(t-t')  f_y(t') dt',  \label{solX}  \ea
\ba
&& Y(t)=Y_0[\chi_0(t)+\Omega^2\omega^2 \chi_2(t)] + X_0 \Omega\omega^2 H_1(t)\cr\cr
&+&V_{y0} [H_0(t)-\Omega^2{\mathcal H}_2(t)] -V_{x0}\Omega {\mathcal H}_1(t) \cr\cr
&+& \int_0^t H_0(t-t') f_y(t') dt'-\Omega^2 \int_0^t {\mathcal H}_2(t-t')  f_y(t') dt' \cr\cr
&-&\Omega \int_0^t {\mathcal H}_1(t-t')  f_x(t') dt',  \label{solY}  \ea
\be
Z(t)=Z_0\chi_0(t)+V_{z0} H_0(t) 
+\int_0^t H_0(t-t') f_z(t') dt',  \label{solZ} \ee
where $X_0=x_0-\langle x_0\rangle$, $Y_0=y_0-\langle y_0\rangle$, and $Z_0=z_0-\langle z_0\rangle$.
The velocities $V_x=v_x-\langle v_x\rangle$, $V_y=v_y-\langle v_y\rangle$, and $V_z=v_z-\langle v_z\rangle$
therefore satisfy
\ba
&&V_x(t)=X_0[-\omega^2H_0(t)+\Omega^2\omega^2 {\mathcal H}_2(t)] 
- Y_0 \Omega\omega^2 {\dot H}_1(t) \cr\cr
&+&V_{x0} [{\dot H}_0(t)-\Omega^2 \dot{\mathcal H}_2(t)] +V_{y0}\Omega \dot{\mathcal H}_1(t) \cr\cr
&+& \int_0^t  {\dot H}_0(t-t') f_x(t') dt' -\Omega^2 \int_0^t \dot{\mathcal H}_2(t-t') f_x(t') dt'  \cr\cr
&+& \Omega \int_0^t \dot{\mathcal H}_1(t-t')  f_y(t') dt' , \label{solVx} \ea
\ba
&&V_y(t)=Y_0[-\omega^2H_0(t)+\Omega^2\omega^2 {\mathcal H}_2(t)] 
+ X_0 \Omega\omega^2 {\dot H}_1(t) \cr\cr
&+&V_{y0} [{\dot H}_0(t)-\Omega^2 \dot{\mathcal H}_2(t)] -V_{x0}\Omega \dot{\mathcal H}_1(t) \cr\cr
&+& \int_0^t {\dot H}_0(t-t') f_y(t') dt' -\Omega^2 \int_0^t \dot{\mathcal H}_2(t-t')  f_y(t') dt' \cr\cr
&-& \Omega \int_0^t \dot{\mathcal H}_1(t-t')  f_x(t') dt' , \label{solVy}  \ea
\be
V_z(t)=-Z_0\omega^2 H_0(t) +V_{z0} {\dot H}_0(t) +\int_0^t {\dot H}_0(t-t') f_z(t') dt',  \label{solVz} \ee
where $V_{x0}=v_{x0}-\langle v_{x0}\rangle$, $V_{y0}=v_{y0}-\langle v_{y0}\rangle$, and
$V_{z0}=v_{z0}-\langle v_{z0}\rangle$. In the absence of the magnetic field ($\Omega=0$), the three
equations (\ref{ddotx}), (\ref{ddoty}), and (\ref{ddotz}) reduce to identical decoupled forms.

The response functions satisfy the initial-value identities
\[
\chi_0(0)=1,\quad \chi_2(0)=0,\quad
H_0(0)=H_1(0)={\mathcal H}_1(0)={\mathcal H}_2(0)=0,
\]
and
\[
\dot H_0(0)=1,\qquad
\dot H_1(0)=\dot{\mathcal H}_1(0)=\dot{\mathcal H}_2(0)=0.
\]
These identities ensure that the explicit solutions above reproduce the prescribed initial positions and velocities at $t=0$.

\section{Gaussian density reduction and positional marginal}
\label{app:pspd}

This appendix gives the Gaussian density reduction used in Sec.~\ref{sec:3}. Because the variables $(\widetilde{\bf R},\widetilde{\bf S})=(X,Y,V_x,V_y)$ are linear functionals of Gaussian initial conditions and Gaussian colored noise, their fixed-time distribution is Gaussian. We write
\be
P(\widetilde{\bf R},\widetilde{\bf S})=
{1\over 4\pi^2\sqrt{\det{\boldsymbol\sigma}(t)}}
\exp\left[-{1\over 2}\tilde{\bf c}^{\,T}{\boldsymbol\sigma}^{-1}(t)\tilde{\bf c}\right],
\label{pspd1}
\ee
where $\tilde{\bf c}=(X,Y,V_x,V_y)^T$. The planar symmetry imposed by the transverse magnetic coupling gives the covariance matrix
\be
{\boldsymbol\sigma}(t)=
\begin{pmatrix}
F&0&H&{\mathcal I}\\
0&F&-{\mathcal I}&H\\
H&-{\mathcal I}&G&0\\
{\mathcal I}&H&0&G
\end{pmatrix},
\label{pssigma}
\ee
with
\[
F=\langle X^2(t)\rangle=\langle Y^2(t)\rangle,\quad G=\langle V_x^2(t)\rangle=\langle V_y^2(t)\rangle,
\]
\[
H=\langle X(t)V_x(t)\rangle=\langle Y(t)V_y(t)\rangle, \quad {\mathcal I}=\langle X(t)V_y(t)\rangle=-\langle Y(t)V_x(t)\rangle .
\]
For the equilibrium initial preparation used in the main text,
\[
\langle X_0^2\rangle=\langle Y_0^2\rangle={k_BT\over k},\quad
\langle V_{x0}^2\rangle=\langle V_{y0}^2\rangle={k_BT\over m},
\]
and the fluctuation--dissipation relation fixes the noise covariance. Substitution of the explicit fluctuation solutions in \ref{app:explicit-solution} gives
\[
F={k_BT\over k},
\qquad
G={k_BT\over m},
\qquad
H=0,
\qquad
{\mathcal I}=0.
\]
Thus the planar phase-space density reduces to
\be
P(\widetilde{\bf R},\widetilde{\bf S})=
{km\over (2\pi k_BT)^2}
\exp\left[-{k|\widetilde{\bf R}|^2+m|\widetilde{\bf S}|^2\over 2k_BT}\right].
\label{pspd-reduced}
\ee
Integrating over the planar velocity fluctuations gives the positional marginal
\be
P(\tilde{\bf r},t)\equiv P(\widetilde{\bf R})=
{k\over 2\pi k_BT}\,
\exp\left[-{k|\widetilde{\bf R}|^2\over 2k_BT}\right],
\label{PR}
\ee
where
\[
|\widetilde{\bf R}|^2=X^2+Y^2=(x-\langle x\rangle)^2+(y-\langle y\rangle)^2.
\]
The corresponding planar positional covariance tensor is
\be
{\boldsymbol\Xi}(t)=\langle\widetilde{\bf R}(t)\widetilde{\bf R}(t)\rangle
={k_BT\over k}\,{\bf I}_2 .
\label{Xi}
\ee
When the Boltzmann constant is absorbed into the temperature, this becomes ${\boldsymbol\Xi}(t)=(T/k){\bf I}_2$. The $z$ sector follows from the same construction by setting $\Omega=0$; it decouples from the planar entropy-production calculation.

\section{Covariance identities for the variance calculations}
\label{app:work-details}

This appendix records the covariance identities used to obtain the work and entropy-production variances in Sec.~\ref{sec:3}. For the direct-force protocol, substituting $x=X+\langle x\rangle$ and $y=Y+\langle y\rangle$ into Eq.~(\ref{Wxy1}) gives the work variance
\be
\sigma_W^2=\int_0^\tau dt\int_0^\tau dt'\,
\dot{\widetilde{\bf F}}_e(t)\cdot
\langle\widetilde{\bf R}(t)\widetilde{\bf R}(t')\rangle\cdot
\dot{\widetilde{\bf F}}_e(t') .
\label{app-varW-cov}
\ee
The fluctuation equations (\ref{ddotX})--(\ref{ddotY}) are autonomous, so equilibrium two-time correlations depend only on time differences. For $t\geq t'$, Eqs.~(\ref{solX}) and (\ref{solY}) give
\ba
\langle X(t)X(t')\rangle&=&\langle Y(t)Y(t')\rangle\nonumber\\
&=&{k_BT\over k}\left[\chi_0(t-t')+\Omega^2\omega^2\chi_2(t-t')\right],
\label{app-XX}
\ea
and
\be
\langle X(t)Y(t')\rangle=-\langle Y(t)X(t')\rangle
={k_BT\over k}\,\Omega\omega^2 H_1(t-t').
\label{app-XY}
\ee
Using these two-time correlations in Eq.~(\ref{app-varW-cov}), integrating by parts, and taking the zero-initial-mean preparation used in Sec.~\ref{sec:3} gives
\ba
\sigma_W^2&=&{2\over\beta}
\left[-\int_0^\tau\dot{\widetilde{\bf F}}_e(t)\cdot
\langle\tilde{\bf r}(t)\rangle\,dt
+{1\over 2m\omega^2}|\widetilde{\bf F}_e(\tau)|^2\right]
\nonumber\\
&=&{2\over\beta}\left[\langle W\rangle
+{1\over 2k}|\widetilde{\bf F}_e(\tau)|^2\right],
\label{app-varW-final}
\ea
which is the identity used in Eq.~(\ref{varW5}).

The entropy-production variance also requires the mixed covariance between the work and the final positional fluctuation. Since $\langle\widetilde{\bf R}(\tau)\rangle={\bf 0}$,
\ba
\langle W\widetilde{\bf R}(\tau)\rangle
&=&-\int_0^\tau
\dot{\widetilde{\bf F}}_e(t)\cdot
\langle\widetilde{\bf R}(\tau)\widetilde{\bf R}(t)\rangle\,dt
\nonumber\\
&=&-{T\over k}\left[\widetilde{\bf F}_e(\tau)-k\langle\tilde{\bf r}(\tau)\rangle\right],
\label{app-WR}
\ea
where the second line uses the convention adopted in Sec.~\ref{sec:3}, in which $k_B$ is absorbed into $T$. Together with
\[
{\boldsymbol\Xi}(\tau)=\langle\widetilde{\bf R}(\tau)\widetilde{\bf R}(\tau)\rangle={T\over k}{\bf I}_2,
\]
Eq.~(\ref{app-WR}) yields
\be
\sigma_s^2=2\langle\Delta s_{\rm tot}\rangle
\ee
for the direct-force protocol.

For the dragged-trap protocol, the fluctuating part of the work is
\[
\hat W-\langle\hat W\rangle
=-\int_0^\tau
\dot{\widetilde{\bf F}}_e(t)\cdot\widetilde{\bf R}(t)\,dt .
\]
It is therefore governed by the same covariance functional as the direct-force work. Consequently,
\be
\sigma_{\hat W}^2={2\over\beta}\langle\hat W\rangle,
\ee
and
\be
\langle\hat W\widetilde{\bf R}(\tau)\rangle
=-{T\over k}\left[\widetilde{\bf F}_e(\tau)-k\langle\tilde{\bf r}(\tau)\rangle\right].
\ee
Substitution into the dragged-trap entropy-production expression gives
\be
\sigma_{\hat s}^2=2\langle\Delta\hat s_{\rm tot}\rangle .
\ee
Thus both protocols use the same covariance mechanism: the linear Gaussian fluctuation dynamics fixes the work covariance, and the equilibrium positional covariance fixes the endpoint entropy contribution.

\end{document}